\documentclass{amsart}
\usepackage{cite,amssymb,amsfonts,amsmath,bm,a4wide,graphicx,amsthm,url}

\renewcommand{\v}[1]{\mathbf{\bm #1}}

\renewcommand{\det}[1]{\left| #1 \right|}

\newcommand{\trace}[1]{\left\langle #1 \right\rangle}

\makeatletter
\g@addto@macro{\endabstract}{\@setabstract}
\newcommand{\authorfootnotes}{\renewcommand\thefootnote{\@fnsymbol\c@footnote}}%
\makeatother

\begin{document}

\begin{center}
  \LARGE
  Efficient sampling of Gaussian graphical models using conditional Bayes factors \par \bigskip

  \normalsize
  \authorfootnotes
  Max Hinne\footnote{\url{mhinne@cs.ru.nl}}\textsuperscript{1}, Alex Lenkoski\textsuperscript{2},
  Tom Heskes\textsuperscript{1}and Marcel van Gerven\textsuperscript{1} \par \bigskip

  \textsuperscript{1}Radboud University Nijmegen, the Netherlands \par
  \textsuperscript{2}Norwegian Computing Center, Norway\par \bigskip

  \today
\end{center}

\begin{abstract}
    Bayesian estimation of Gaussian graphical models has proven to be challenging because the conjugate prior distribution on the Gaussian precision matrix, the $G$-Wishart distribution, has a doubly intractable partition function. Recent developments provide a direct way to sample from the $G$-Wishart distribution, which allows for more efficient algorithms for model selection than previously possible. Still, estimating Gaussian graphical models with more than a handful of variables remains a nearly infeasible task. Here, we propose two novel algorithms that use the direct sampler to more efficiently approximate the posterior distribution of the Gaussian graphical model. The first algorithm uses conditional Bayes factors to compare models in a Metropolis-Hastings framework. The second algorithm is based on a continuous time Markov process. We show that both algorithms are substantially faster than state-of-the-art alternatives. Finally, we show how the algorithms may be used to simultaneously estimate both structural and functional connectivity between subcortical brain regions using resting-state fMRI.
\end{abstract}

\section{Introduction}
A key objective in many areas of science is to uncover the interactions amongst a large number of variables based on a limited amount of data. Examples include gene regulatory networks, where one wants to identify the interactions amongst DNA segments, market basket analysis where the relations are studied between customers based on their purchase behavior, or neuroscience where the connections between segregated neuronal populations are linked to cognitive ability and impairment. One way to estimate these relations is to employ Gaussian graphical models, where the non-zero entries in the off-diagonal of a precision matrix correspond to the edges in a conditional independence graph~\cite{Dempster1972}. However, fully Bayesian estimation of the posterior of a Gaussian graphical model has proven to be notoriously hard.

To allow Bayesian inference of the Gaussian graphical model, a conjugate prior~\cite{Diaconis1979} on a precision matrix restricted by the conditional independence graph $G$ was constructed for decomposable graphs~\cite{Dawid1993}, and later generalized to arbitrary graphs~\cite{Roverato2002}. Subsequent work coined this distribution the $G$-Wishart distribution~\cite{AtayKayis2005}. A number of Monte Carlo algorithms for model estimation using the $G$-Wishart distribution have been developed~\cite{Piccioni2000,Mitsakakis2011,Dobra2011,Wang2012}, but each of these algorithms required substantial computational resources due to difficulty with sampling from the $G$-Wishart distribution. To address this bottleneck, a recent study proposed an efficient way to directly sample from the $G$-Wishart distribution~\cite{Lenkoski2013} by scaling samples from a regular Wishart distribution to fit the required dependency structure~\cite{Hastie2009}. Even with the direct sampler, approximating the Gaussian graphical model remained difficult because of the doubly intractable partition function of the $G$-Wishart distribution. However, by combining features of the exchange algorithm~\cite{Murray2006} with reversible jump sampling~\cite{Green1995}, calculating the partition function may be circumvented~\cite{Lenkoski2013}. The algorithm that implements this idea, named the double reversible jump algorithm, provides substantial computational gains compared to earlier approaches~\cite{Lenkoski2013}.

Although the double reversible jump algorithm enables model selection in a more efficient manner than previously possible, computational costs remain a limiting factor in practical applications with a large number of variables. In this paper, we propose two novel, faster, algorithms for Bayesian estimation of the Gaussian graphical model. In the first algorithm, we combine the direct sampler~\cite{Lenkoski2013} with an efficient representation of the conditional Bayes factor~\cite{Cheng2012}, which results in an elegant Metropolis-Hastings algorithm to which we will refer as the double conditional Bayes factor sampler. In the second algorithm, we cast the double conditional Bayes factors algorithm in a birth-death MCMC setting~\cite{Mohammadi2014}. Here, rather than accepting or rejecting a new state with an edge added or removed, we associate with these changes birth and death events, respectively. These events occur with such rates that their equilibrium coincides with the posterior of interest~\cite{Stephens2000}. Both algorithms provide substantial speed improvement over the status quo, as we show in simulations.

We also provide an application of our algorithms by estimating structural and functional connectivity between subcortical structures using resting-state fMRI. It is a major goal in cognitive neuroscience to understand how spatially segregated neural populations are coupled, using indirect measures of neural activity such as functional magnetic resonance imaging~\cite{Smith2013,Salinas2001}. In this context, the anatomical pathways between neural populations are referred to as structural connectivity whereas correlated activity patterns between these populations are referred to as functional connectivity~\cite{Friston2011}. Both forms of connectivity may be estimated simultaneously using Gaussian graphical models. Here, the precision matrix captures the functional interactions between variables and the associated conditional independence graph represents the direct connections between variables. Bayesian estimation of Gaussian graphical models is particularly relevant since the posterior over precision matrices provides complete information about the strength of functional interactions and the posterior over conditional independence graphs allows one to associate a probability with a putative direct connection between variables of interest.

\section{Gaussian graphical models}
\subsection{Preliminaries}

Let observed data $\v{X}=(\v{x}_1, \ldots, \v{x}_n)^T$ consist of $n$ independent draws from a $p$-dimensional multivariate Gaussian distribution $\mathcal{N}(\v{0},\v{K}^{-1})$, with zero mean and precision (inverse covariance) matrix $\v{K}$. Here, $\v{K}\in \mathbb{P}_p$, with $\mathbb{P}_p$ the space of positive definite $p\times p$ matrices. The likelihood of $\v{K}$ is given by
\begin{equation}\label{eq:likelihood}
    P(\v{X}\mid \v{K}) = \prod^n_{i=1} \mathcal{N}(\v{x}_i \mid \v{0},\v{K}^{-1}) \propto \det{\v{K}}^{n/2} \exp\left[-\frac{1}{2} \trace{\v{K},\v{S}} \right] \enspace,
\end{equation} where $\v{S}=\v{X}^T\v{X}$ is the empirical covariance and $\trace{\cdot,\cdot}$ the trace inner product operator.
The precision matrix has the important property that zero elements correspond to conditional independencies. In other words,~\eqref{eq:likelihood} specifies a Gaussian Markov random field with respect to a graph $G=(V,E)$, with $V=\{1,\ldots,p\}$ and $E\subset V\times V$, in which the absence of a connection indicates independence, i.e. $(i,j)\not\in E \rightarrow k_{ij}=0$. For convenience, throughout this paper we slightly abuse notation and use $(i,j)\in G$ to indicate that the edge $(i,j)$ is present in $E$.

The dependency graph may be used to specify a prior distribution on the precision matrix, which is known as the $G$-Wishart distribution~\cite{Roverato2002}:
\begin{equation}
    P(\v{K}\mid G, \delta, \v{D}) = \mathcal{W}_G(\delta,\v{D}) = \frac{\det{\v{K}}^{(\delta-2)/2}}{Z_G(\delta, \v{D})} \exp\left[ -\frac{1}{2} \trace{\v{K},\v{D}} \right] \v{1}_{\v{K}\in\mathbb{P}_G} \enspace,
\end{equation} in which $\mathbb{P}_G$ is the space of positive definite $p\times p$ matrices that have zero elements wherever $(i,j)\not\in G$, $\delta$ is the prior degrees of freedom, $\v{D}$ is the prior scaling matrix and $\v{1}_x$ evaluates to 1 if and only if $x$ holds and to 0 otherwise. The $G$-Wishart distribution is conjugate to the multivariate Gaussian likelihood in~\eqref{eq:likelihood}, so that
\begin{equation}
    P(\v{K}\mid G, \delta, \v{D}, \v{X}) = \mathcal{W}_G(\delta+n, \v{D}+\v{S}) = \frac{\det{\v{K}}^{(n+\delta-2)/2}}{Z_G(\delta+n, \v{D}+\v{S})} \exp\left[ -\frac{1}{2} \trace{\v{K},\v{D}+\v{S}} \right] \enspace.
\end{equation} Note that the Wishart distribution is a special case of the $G$-Wishart distribution, with which it coincides if $G$ is a fully connected graph. Importantly, the partition function $Z_G(\delta, \v{D})$ depends on $G$, which makes the $G$-Wishart a doubly intractable distribution. We return to the implications of this fact later on.

Central to this work is that we wish to perform model selection in Gaussian graphical models, which revolves around the joint posterior
\begin{equation}\label{eq:posterior}
    P(G,\v{K} \mid \v{X}) \propto P(\v{X}\mid \v{K}) P(\v{K} \mid G) P(G) \enspace.
\end{equation} In the remainder, we outline several algorithms to approximate this distribution.

\subsection{Direct samples from the $G$-Wishart distribution}\label{sec:sampler}
Since the prior $P(\v{K}\mid G)$ is $\mathcal{W}_G(\delta,\v{D})$, we need a way to draw samples from the $G$-Wishart distribution. Up until recently, this was achieved using a block Gibbs sampler that updates $\v{K}$ according to either the edges of $G$~\cite{Wang2012} or its clique decomposition~\cite{Piccioni2000}. Although this enables model inference of $P(G, \v{K}\mid \v{X})$, as desired, both approaches require substantial computational effort, making them prohibitive for use in contexts with a large number of variables. An alternative method was proposed that is more efficient~\cite{Lenkoski2013}, which is an adaptation of an algorithm for estimating the mode $\hat{\v{K}}$ of the $G$-Wishart distribution~\cite{Hastie2009,Moghaddam2009}. The algorithm is as follows:

\begin{enumerate}
    \item Sample $\v{\Sigma} \sim \mathcal{W}(\delta,\v{D})$.
    \item Let $\v{W} = \v{\Sigma}$.
    \item Repeat for $j=1,2,\ldots,p$ until convergence:
    \begin{enumerate}
        \item Let $N_j \subset V$ be the set of variables that are connected to $j$ in $G$.
        \\Form $\v{W}_{N_j}$ and $\v{\Sigma}_{N_j,j}$ and solve
            \begin{equation*}
                \hat{\beta}^*_j = \v{W}^{-1}_{N_j}\v{\Sigma}_{N_j,j}
            \end{equation*}
        \item Form $\hat{\beta}_j \in \mathbb{R}^{p-1}$ by copying the elements of $\hat{\beta}^*_j$ to the appropriate \\locations and imputing zeros in those locations not connected to $j$ in $G$.
        \item Replace $\v{W}_{j,-j}$ and $\v{W}_{-j,j}$ with $\v{W}_{-j,-j}\hat{\beta}_j$.
    \end{enumerate}
    \item Return $\v{K} = \v{W}^{-1}$.
\end{enumerate} Conceptually, the algorithm draws a sample from a Wishart distribution, which is then iteratively scaled according to the dependence structure in $G$. In practice, we observe that convergence (see step 3) is typically reached within a handful of iterations, even for moderate to large $p$.

\section{Sampling algorithms}
The direct sampler paves the way for novel inference algorithms. Here, we introduce two novel algorithms for approximation of the joint posterior in~\eqref{eq:posterior}.

\subsection{Double reversible jump sampler}
As a baseline for comparison, we use the double reversible jump (DRJ) sampler~\cite{Lenkoski2013}. This algorithm was shown to be more efficient as previously used approaches and may be considered  state of the art. It builds upon the reversible jump sampler discussed in~\cite{Dobra2011a}. The key idea offered by this approach is that it introduces an auxiliary variable $\v{K}^0\sim\mathcal{W}_G(\delta,\v{D})$, as in the exchange algorithm~\cite{Murray2006}, that is efficiently sampled using the direct sampler discussed above. Because of the way this auxiliary variable is constructed, the doubly intractable partition functions of the $G$-Wishart distribution are canceled out in the calculation of the acceptance ratios of newly proposed graphs.

\subsection{Direct double conditional Bayes factor sampler}
The double reversible jump algorithm provides a substantial improvement over previous algorithms, as it avoids the need to approximate the ratio of partition functions or invoke the Gibbs sampling algorithm for drawing samples from the $G$-Wishart distribution. Nonetheless, the algorithm can be simplified. In~\cite{Cheng2012} it is shown that if $G$ and $\tilde{G}$ differ only in the edge $e=(p-1,p)$ and $G\subset \tilde{G}$, the odds ratio of these two models may be expressed as
\begin{equation}\label{eq:cbf}
    \frac{P(\v{X}\mid \tilde{G}, \v{K}, \v{D})}{P(\v{X}\mid G, \v{K}, \v{D})} = N(\v{K}, \v{D} + \v{S}) \frac{Z_G(\delta,\v{D})}{Z_{\tilde{G}}(\delta,\v{D})} 
\end{equation} with
\begin{equation}
    N(\v{K}, \v{U}) \equiv \phi_{p-1,p-1} \left( \frac{2\pi}{u_{pp}} \right)^{1/2} \exp \left[ \frac{1}{2} u_{pp} \left( \frac{\phi_{p-1,p-1}u_{p-1,p}}{u_{pp}} - \frac{\sum_{l=1}^{p-2} \phi_{lp-1}\phi_{lp}}{\phi_{p-1,p-1}} \right)^2 \right] \enspace, 
\end{equation} where $\v{K}=\Phi^T\Phi$, with $\Phi$ the Cholesky decomposition of $\v{K}$. The term in~\eqref{eq:cbf} can be considered the conditional Bayes factor of the comparison between $G$ and $\tilde{G}$.
Similar to the double reversible jump approach,~\cite{Cheng2012} propose to augment the sampling process with an auxiliary variable $\v{K}^0\sim\mathcal{W}_G(\delta,\v{D})$. This results in a convenient acceptance ratio for the addition of an edge to $G$
\begin{equation}\label{eq:cbf_ar}
    \alpha = \frac{N(\v{K}, \v{D}+\v{S})}{N(\tilde{\v{K}}^0, \v{D})}\frac{P(\tilde{G})}{P(G)} \enspace,
\end{equation} where the ratio is inverted if the edge is removed from $G$ instead. Note that the variables $G, \v{K}, \v{U}$ and $\v{D}$ must be permuted for each edge flip to place the particular edge under consideration in the position $(p-1,p)$.

The algorithm described in~\cite{Cheng2012} employs the block Gibbs sampler to sample from the $G$-Wishart distribution. Instead, here we propose to make use of the direct sampler explained in Section~\ref{sec:sampler} to arrive at the following procedure for estimation of the Gaussian graphical model:
\begin{enumerate}
    \item Let $G=G^{[s]}$ be the current graph and let $\v{K}=\v{K}^{[s]}\sim \mathcal{W}_G(\delta+n, \v{D}+\v{S})$.
    \item For each edge $(i,j)\in G$, do:
    \begin{enumerate}
        \item Create a permutation of the variables so that $(i,j)\rightarrow (p-1,p)$. Permute $G,\v{K},\v{D}$ and $\v{S}$ accordingly.
        \item Let $\tilde{G}=G\cup (p-1,p)$ if $(p-1,p)\not\in G$ or $\tilde{G}=G\setminus (p-1,p)$ if $(p-1,p)\in G$.
        \item Draw $\tilde{\v{K}}^0\sim \mathcal{W}_{\tilde{G}}(\delta,\v{D})$.
        \item Accept the move from $G$ to $\tilde{G}$ with probability $\alpha$ as in~\eqref{eq:cbf_ar}.
        \item Restore the original ordering of $G,\v{K},\v{D}$ and $\v{S}$ and draw $\tilde{\v{K}}\sim\mathcal{W}_{\tilde{G}}(\delta+n, \v{D}+\v{S})$
    \end{enumerate}
    \item Set $G^{[s+1]}=\tilde{G}$ and $\v{K}^{[s+1]}\sim \mathcal{W}_{\tilde{G}}(\delta+n, \v{D}+\v{S})$.
\end{enumerate}

The usage of the direct sampler instead of the block Gibbs updates makes this direct double conditional Bayes factors (DCBF) algorithm computationally much more efficient~\cite{Liang2010}.


\subsection{Double continuous time sampler}
A downside of the usage of an auxiliary variable scheme is that it decreases the acceptance probability of proposals, as essentially two moves have to be accepted at once. This hampers mixing of the Markov chain, so that multimodal distributions are approximated poorly. To improve acceptance,~\cite{Mohammadi2014} introduce a birth-death continuous-time Markov process~\cite{Cappe2003} for Gaussian graphical models. Rather than accepting the addition or removal of an edge,~\cite{Mohammadi2014} associates birth and death events with these changes, respectively. Each edge dies independently of all others as a Poisson process with death rate $d_e(G,\v{K})$. Because the edges are independent, the overall death rate at a particular pair of graph $G$ and precision $\v{K}$ is $d(\v{K})=\sum_e d_e(G,\v{K})$. Birth rates $b(\v{K})$ are defined similarly, but for non-edges instead.

Because the birth and death processes are independent Poisson processes, the expected time between two events is $1/(d(\v{K})+b(\v{K}))$. This time can be considered the process spends at any particular instance of $(G,\v{K})$. The probability of the death event of edge $e\in G$ is
\begin{equation}\label{eq:probdeath}
    P(\mbox{death of edge } e) = \frac{d(G,\v{K})}{b(G,\v{K})+d(G,\v{K})}  \enspace,
\end{equation} with again an analogous definition for the birth event for a non-edge.

Mohammadi and Wit~\cite{Mohammadi2014} show that the birth-death process has the posterior $P(G,\v{K}\mid \v{X})$ as stationary distribution, if for all edges and non-edges $e$
\begin{equation}
    d_e(\tilde{G},\tilde{\v{K}}) P(\tilde{G},\tilde{\v{K}}\mid \v{X}) = b_e(G,\v{K}) P(G,\v{K}\mid \v{X}) \enspace,
\end{equation} for $\tilde{G}=G\cup e$. The birth and death rates may be chosen accordingly as
\begin{equation}\label{eq:birthdeath}
    b_e(G,\v{K}) = \frac{P(\tilde{G},\tilde{\v{K}}\mid \v{X})}{P(G,\v{K}\mid \v{X})}  \quad \mbox{for $e\not\in G$} \quad \mbox{and} \quad d_e(G,\v{K}) = \frac{P(G,\v{K}\mid \v{X})}{P(\tilde{G},\tilde{\v{K}}\mid \v{X})}  \quad \mbox{for $e\in G$}.
\end{equation} with again $\tilde{G}=G\cup e$.

The key observation is now that these birth-death rates can be computed using the double conditional Bayes factors as in~\eqref{eq:cbf_ar}. Here again we make use of the exchange framework by introducing the auxiliary variable $\v{K}^0$, such that explicit evaluation of the partition functions is circumvented. This leads to a novel approach that we will refer to as the double continuous time (DCT) sampler, given by:
\begin{enumerate}
    \item Let $G=G^{[s]}$ be the current graph and let $\v{K}=\v{K}^{[s]}\sim \mathcal{W}_G(\delta+n, \v{D}+\v{S})$.
    \item For each non-edge $e\not\in G$:
    \begin{enumerate}
        \item Create a random permutation of the variables so that $(i,j)\rightarrow (p-1,p)$. Permute $G,\v{K},\v{D}$ and $\v{S}$ accordingly.
        \item Let $\tilde{G}=G\cup e$. Draw $\v{K}^0\sim \mathcal{W}_{\tilde{G}}(\delta,\v{D})$
        \item Compute the birth rate $b_e(G,\v{K})$ using~\eqref{eq:birthdeath}.
    \end{enumerate}
    \item Compute the total birth rate of the current state $b(G,\v{K})$.
    \item For each edge $e\in G$:
    \begin{enumerate}
        \item Create a random permutation of the variables so that $(i,j)\rightarrow (p-1,p)$. Permute $G,\v{K},\v{D}$ and $\v{S}$ accordingly.
        \item Let $\tilde{G}=G\setminus e$. Draw $\v{K}^0\sim \mathcal{W}_{\tilde{G}}(\delta,\v{D})$
        \item Compute the death rate $d_e(G,\v{K})$ using~\eqref{eq:birthdeath}.
    \end{enumerate}
    \item Compute the total death rate of the current state $d(G,\v{K})$ and the waiting time between events $w(G,\v{K}) = 1/(d(\v{K})+b(\v{K}))$.
    \item Create a birth or death event according to the probabilities of death events~\eqref{eq:probdeath} and birth events, and set $G^{[s+1]}=\tilde{G}$ and $\v{K}^{[s+1]} \sim \mathcal{W}_{\tilde{G}}(\delta+n, \v{D}+\v{S})$.
\end{enumerate}

\section{Experiments}
In this section we first analyze the validity of the two proposed methods using an example with a known precision matrix. Subsequently we apply the algorithms in an explorative study to identify structural and functional connectivity between subcortical brain structures.

\subsection{Simulation}
We compared the performance of the double reversible jump algorithm and the two novel algorithms using a simulation proposed in~\cite{Wang2012}. In this example, we have $p=6$ and $n=18$. Furthermore, the precision matrix $\v{K}$ is given by $k_{ii}=1$ for $i=1,\ldots,p$, $k_{i,i+1}=k_{i+1,i}=0.5$ for $i=1,\ldots,p-1$ and finally $k_{1p}=k_{p1}=0.4$. The associated conditional independence graph $G$ follows as $(i,j)\in G \leftrightarrow k_{ij}\neq 0$. The scatter matrix is then constructed as $\v{S}=\v{X}\v{X}^T =n\v{K}^{-1}$, which corresponds to $n$ independent observations of $\mathcal{N}(\v{0}, \v{K}^{-1})$. Through exhaustive enumeration of all 32,768 possible graphs of size $p$,~\cite{Wang2012} shows that the posterior edge probabilities are
\begin{equation}\label{eq:edges}
\begin{small}
    P((i,j)\in G \mid \v{X}) = \left( \begin{array}{rrrrrrr} 1& 0.969 & 0.106& 0.085 & 0.113 & 0.850\\ 0.969 & 1 & 0.980 & 0.098 & 0.081 & 0.115 \\ 0.106 & 0.980 & 1 & 0.982 & 0.098 & 0.086\\ 0.085 & 0.098 & 0.982 & 1 & 0.980 & 0.106 \\ 0.113 & 0.081 & 0.98 & 0.980 & 1 & 0.970\\ 0.850 & 0.115 & 0.086 & 0.106 & 0.970 & 1 \end{array}  \right)
    \end{small}
\end{equation} and the expectation of $\v{K}$ is
\begin{equation}\label{eq:prec}
\begin{small}
    \mathbb{E}(\v{K} \mid \v{X}) = \left( \begin{array}{rrrrrrr} 1.139& 0.569 & -0.011& 0.006 & -0.013 & 0.403\\ 0.569 & 1.175 & 0.574 & -0.008 & 0.005 & -0.014 \\ -0.011 & 0.574 & 1.176 & 0.574 & -0.008 & 0.006\\ 0.006 & -0.008 & 0.574 & 1.175 & 0.573 & -0.011 \\ -0.013 & 0.005 & -0.008 & 0.573 & 1.175 & 0.569\\ 0.403 & -0.014 & 0.006 & -0.011 & 0.569 & 1.138 \end{array}  \right) \enspace.
\end{small}
\end{equation} We approximate this ground truth using the three different algorithms, each implemented in Matlab. Throughout, we use vague priors in the form of $P(G)\propto 1$ for $G$ and $P(\v{K}\mid G)=\mathcal{W}_G(3,\v{I}_p)$. The algorithms are each executed for 100,000 iterations, of which the first 50,000 are discarded as burn-in. Conditional expectations for edges (i.e.~edge probabilities) and precision matrices are then calculated as \begin{equation}
    \mathbb{E}((i,j) \in G \mid \v{X}) = \frac{1}{T}\sum_{t=1}^T \v{1}_{(i,j)\in G_t}\quad\mbox{and}\quad\mathbb{E}(\v{K} \mid \v{X}) = \frac{1}{T}\sum_{t=1}^T \v{K}_t
\end{equation} for the double reversible jump and the double conditional Bayes factor algorithms, with $T$ the number of samples. For the double continuous time algorithm, these expectations are calculated as
\begin{equation}\label{eq:ctexp}
    \mathbb{E}((i,j)\in G \mid \v{X}) = \frac{1}{W} \sum_{t=1}^T w_t \v{1}_{(i,j)\in G_t} \quad\mbox{and}\quad \mathbb{E}(\v{K} \mid \v{X}) = \frac{1}{W} \sum_{t=1}^T w_t \v{K}_t \enspace,
\end{equation} with $W=\sum_{t=1}^T w_t$. It is easy to see that this idea generalizes the discrete time MCMC approach by assuming $w_t=1$ for all $t$. 

We quantify the approximation accuracy of the three algorithms in a number of ways. First, the accuracy of the edge probabilities is expressed using the mean squared error with respect to the true probabilities in~\eqref{eq:edges}. Second, we compute the Kullback-Leibler divergence~\cite{Kullback1951} between the precision matrix obtained in~\cite{Wang2012} as defined in~\eqref{eq:prec} and $\hat{\v{K}}\equiv\mathbb{E}(\v{K} \mid \v{X})$ using either of the algorithms. We also count the number of unique models that each algorithm considers to express mixing behavior. Next, we compute the marginal posterior probability of the true graph. Finally, we compute the relative computational speeds of the algorithms. The results of the comparison are shown in Table~\ref{tab:simresults}. The algorithms have similar performance in approximating the desired posterior distribution and each obtains the true graph as the mode of the approximated distribution. Contrary to~\cite{Mohammadi2014}, we do not find the continuous time algorithm to have the best mixing. In fact, of the three considered models, the continuous time MCMC approach finds the smallest number of unique models. Note that the continuous time approach may converge faster~\cite{Rao2012}, but this is not apparent in this simulation. Finally, the efficiency of our way of computing the conditional Bayes factor (see~\eqref{eq:cbf}) is demonstrated by a substantial speed increase, as the DCBF algorithm is 3.57 times faster than the DRJ sampler, and the DCT algorithm is 3.80 times faster than the DRJ algorithm, whereas the algorithm in~\cite{Mohammadi2014} is 1.79 times slower than the DRJ sampler.

\begin{table}[t]
    \caption{Results for the comparison between the three described samplers on a simulated example, averaged over 10 simulations. Standard errors are indicated in parentheses. Shown are the mean squared error (MSE) of edge probabilities relative to~\eqref{eq:edges}, the Kullback-Leibler divergence (KL) between the expected precision matrix and~\eqref{eq:prec}, the number of unique models visited, the marginal posterior probability of the true graph $P(G\mid \v{S})$ and the relative speed of the algorithms compared to the double reversible jump baseline.}
    \label{tab:simresults}
    \begin{center}
    \begin{tabular}{lrrrrr}
     \multicolumn{1}{c}{\bf Algorithm} &  \multicolumn{1}{c}{\bf MSE} &  \multicolumn{1}{c}{\bf KL} &  \multicolumn{1}{c}{\bf \#models} & \multicolumn{1}{c}{$P(G\mid\v{S})$} & \multicolumn{1}{c}{\bf Rel. speed} \\
      \hline
      DRJ & 5e-04 (4e-05) & 1e-04 (2e-05)& 1299 (31) &0.3674 (0.0008)& 1 (0) \\
      DCBF & 5e-04 (2e-05) & 1e-04 (1e-05)& 1472 (23) &0.3826 (0.0040)& 3.57 (1e-01) \\
      DCT & 1e-03 (1e-05)& 7e-04 (3e-04)& 1187 (35) &0.4277 (0.0008)& 3.80 (1e-02) \\
      \hline
    \end{tabular}
    \end{center}
\end{table}

\subsection{Subcortical brain connectivity}

As an explorative example, we estimate structural and functional connectivity in a fully Bayesian setting. In previous work, functional connectivity has been estimated under the assumption that the underlying structural connectivity was known~\cite{Hinne2014}. Here, we address the more challenging problem of simultaneously estimating the posterior distribution of both structural and functional connectivity.

\begin{figure}[ht]
  \centering
  \includegraphics[width=\textwidth]{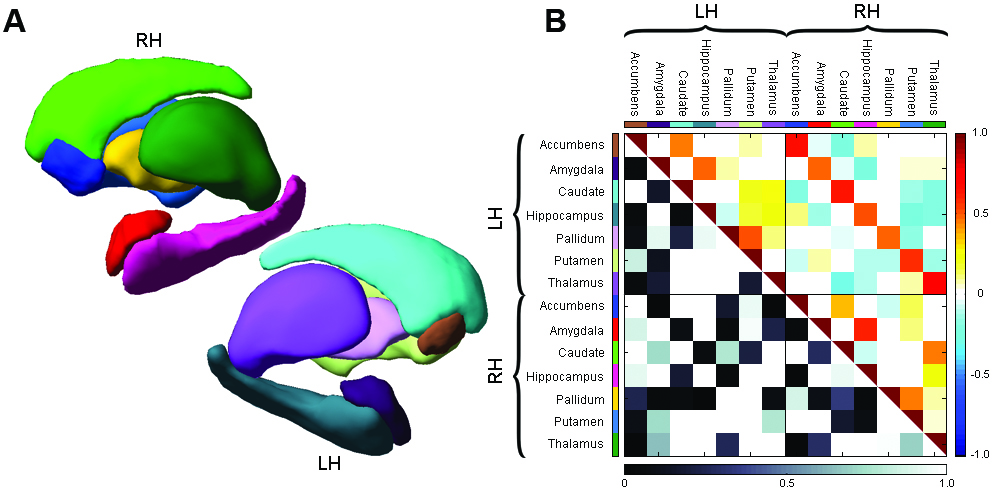}\\
  \caption{Subcortical connectivity. A. Subcortical structures, consisting of bilateral accumbens, amygdala, caudate, hippocampus, pallidum, putamen and thalamus. B. Posterior probabilities of structural connectivity (lower triangle) and expected partial correlations between these structures (upper triangle). LH and RH indicate left hemisphere, right hemisphere, respectively.}\label{fig:subcort}
\end{figure}

\subsubsection{Empirical data}
The data consist of resting-state functional MRI data collected for one subject. We refer the reader to~\cite{VanOort2014} for details of the acquisition protocol. Preprocessing was performed using FSL 5.0~\cite{Jenkinson2012} and consisted of the following steps. T1 images were linearly registered to MNI-152 space. Multi-echo volumes at each TR were combined~\cite{Poser2006}. Motion correction was performed using MCFLIRT and estimated motion parameters were regressed out together with their temporal derivatives and mean time courses for both WM and CSF. Finally, data were high-pass filtered at 0.001 Hz. Subcortical structures were segmented using FSL FIRST~\cite{Patenaude2011}, resulting in data for a total of 14 regions, consisting of bilateral accumbens, amygdala, caudate, hippocampus, pallidum, putamen and thalamus (see Fig.~\ref{fig:subcort}A). For each of these regions the signal was averaged over all voxels in that region and subsequently standardized to have zero mean and unit variance.

\subsubsection{Bayesian structural and functional connectivity estimation}
The human brain can be viewed as a complex dynamical system where ongoing changes in neuronal dynamics produce adaptive behavior~\cite{Bullmore2009}. These dynamics can be expressed in terms of interactions between brain regions, which is commonly referred to as functional connectivity. At the same time, direct functional interactions presuppose anatomical links between brain regions, known as structural connectivity. For this reason, structural and functional connectivity must be intimately related~\cite{Akil:2011kb}.

Functional connectivity is most easily expressed using a covariance matrix that, in the case of standardized data, provides the correlation structure between different brain regions. However, this approach suffers from the drawback that it cannot distinguish between direct and indirect connections. Alternatively, one may use partial correlations that capture only direct effects, in the absence of confounding factors. The matrix of partial correlations $\v{R}$ may be obtained from a precision matrix using $r_{ij}=1$ if $i=j$ and $r_{ij}=-k_{ij}/\sqrt{k_{ii}k_{jj}}$ otherwise. Because functional coupling must be accompanied by an anatomical connection, partial correlations between brain regions not only reveal the strength of these couplings, but also indicate which regions are physically connected. In other words, the joint posterior in~\eqref{eq:posterior} becomes a distribution over functional connectivity $\v{K}$ (or, equivalently, $\v{R}$) and structural connectivity $G$.

We proceed by approximating the joint posterior using both the DCBF algorithm as well as the DCT sampler. Both algorithms were executed for 100,000 iterations, of which the first 50,000 were discarded as burn in. Once again, we set $P(G)\propto1$ and $P(\v{K}\mid G)=\mathcal{W}_G(3,\v{I}_p)$. The algorithms yield almost identical results, as shown by an MSE of edge probabilities of 0.0006 and a symmetrized Kullback-Leibler divergence of 0.0002.

Figure~\ref{fig:subcort}B shows the posterior edge probabilities and partial correlations produced by the DCT algorithm. The structural connectivity estimate shows that the majority of edges is associated with either very high or very low edge probabilities. The functional connectivity estimate shows that functional homologues in left and right hemispheres are associated with high partial correlations (expected partial correlations $\langle r \rangle$ in the range $[0.48,0.73]$), indicating that these functional homologues have similar functional roles. Within a cortical hemisphere, the most salient functional interactions (highest expected partial correlations with $\langle r \rangle$ in the range $[0.23,0.61]$) are given bilaterally by amygdala$-$hippocampus, pallidum$-$putamen, accumbens$-$caudate, caudate$-$thalamus, and hippocampus$-$thalamus. These functional interactions can be explained by direct pathways as well as unobserved common inputs that induce a high partial correlation. Interestingly, edges with high posterior probability (edge probability higher than 0.999) can be associated with weak absolute partial correlations (with $\langle r \rangle$ as low as 0.1). This indicates that there exist weakly coupled regions (from the linear correlation point of view) that cannot be explained away by other functional interactions.

\section{Discussion}

We have proposed two novel algorithms for Bayesian model selection in a Gaussian graphical model. The first algorithm combines a direct manner to sample $G$-Wishart variates~\cite{Lenkoski2013} with an efficient way of computing conditional Bayes factors when comparing two different models~\cite{Cheng2012}, resulting in an improved Metropolis-Hastings approach. The second approach integrates the direct sampler within a birth-death continuous time Markov process~\cite{Mohammadi2014}. Both algorithms provide accurate estimates of the posterior graphs and precision matrices and are substantially faster (up to a factor of 3.80) than previously available alternatives. We demonstrate the use of the algorithms by estimating, for the first time, both structural and functional connectivity simultaneously using fMRI data.

In future work we aim to improve mixing of the samplers by introducing moves between graphs that differ by more than a single edge. Similarly, one may conceive events other than births and deaths of edges. In either case, the corresponding conditional Bayes factors must be derived, and these should be more efficient to compute than a series of consecutive edge additions and removals. We expect that this will further contribute to efficient estimation of Gaussian graphical models.





\subsubsection*{Acknowledgments}

The authors gratefully acknowledge the support of the BrainGain Smart Mix Programme of the Netherlands Ministry of Economic Affairs and the Netherlands Ministry of Education, Culture and Science. Alex Lenkoski's work is supported by Statistics for Innovation, (sfi)$^2$, in Oslo. The authors thank Erik van Oort and David Norris for the acquisition of the fMRI data.

\small{
\bibliographystyle{unsrt}
\bibliography{library}
}

\end{document}